\documentclass[aps,prr,twoside,twocolumn,amsfont,amssymb,amsmath,showpacs,preprintnumbers]{revtex4-1}
\usepackage{breqn}
\usepackage{etoolbox}
\usepackage{amsmath}
\usepackage{bm}
\usepackage{color}
\usepackage{dcolumn}
\usepackage{float}
\usepackage{graphicx}
\usepackage{epstopdf}
\usepackage[colorlinks,linkcolor=red,citecolor=blue]{hyperref}
\usepackage{mathbbol}
\usepackage{mathrsfs}
\usepackage{placeins}
\usepackage{verbatim}
\usepackage{url}
\usepackage{breakurl}

\newcommand{\CAP}{{\scriptscriptstyle\mathrm{CAP}}}
\newcommand{\XUV}{{\scriptscriptstyle\mathrm{XUV}}}

\newcommand{\IR}{{\scriptscriptstyle\mathrm{IR}}}

\newcommand{\sumint}{\sum\hspace{-13pt}\int}

\begin{document}

\title{ Coherence control in Helium-Ion ensembles }

\author{Saad Mehmood$^1$,  Eva Lindroth$^2$, and Luca Argenti$^{1,3}$}\email{luca.argenti@ucf.edu} 
\affiliation{$^1$Department of Physics University of Central Florida}
\affiliation{$^2$Department of Physics Stockholm University Stockholm}
\affiliation{$^3$CREOL University of Central Florida Orlando Florida}


\begin{abstract}
Attosecond pulses can ionize atoms in a coherent process. Since the emerging fragments are entangled, however, each preserves only a fraction of the initial coherence, thus limiting the chance of guiding the ion subsequent evolution. In this work, we use \emph{ab initio} simulations of pump-probe ionization of helium above the $2s/2p$ threshold to demonstrate how this loss of coherence can be controlled. Thanks to the participation of $2\ell n\ell'$ states, coherence between the ionic $2s$ and $2p$ states, which are degenerate in the non-relativistic limit, results in a stationary, delay-dependent electric dipole. From the picosecond real-time beating of the dipole, caused by the fine-structure splitting of the $n=2$ manifold, it is possible to reconstruct all original ion coherences, including between antiparallel-spin states, which are sensitive probe of relativistic effects in attosecond photoemission. 
\end{abstract}

\maketitle

\section{Introduction}
Atomic systems feature excited bound and mestastable states separated in energy by several eV or even tens or hundreds of eV. Coherent superpositions of these states, therefore, give rise to electronic motions that unfold on a sub-femtosecond timescale~\cite{Krausz2009,RevModPhys.87.765,Calegari_2016}. Attosecond XUV-pump IR-probe photoelectron spectroscopies have emerged as powerful tools to explore charge-transfer processes in complex systems~\cite{Haessler2010,Goulielmakis2008, Itatani2002, Villoresi2006,Horvath2007} and attosecond dynamics at the nanoscale ~\cite{Ciappina_2017,DombiNano_2020}. The short duration of attosecond pulses generates coherent superposition of electronic states above the ionization threshold, bearing the promise of quantum control in the electronic continuum~\cite{SFI_Pabst,DecoherencePRX}.  In a photoionization process, however, the photoelectron and its parent ion form an entangled pair. As soon as the photoelectron leaves the interaction region, therefore, part of the coherence in the residual parent-ion is lost, and so is the chance of guiding any subsequent transformation of the target.
One way to limit the loss of coherence that accompanies photoionization is to polarize the target with a strong control field that forces the ion in a single polarized state~\cite{Ossiander2016,GoulielmakisNAT2010}. In a theoretical study of the one-photon ionization of xenon, which can result in partial coherence between ions with the same parity, Pabst \emph{et al.} have shown this coherence increasing for pulses with shorter duration and higher central frequency, on account of the reduced role of inter-channel coupling at large photoelectron energies~\cite{Pabst2011,PabstPRA}.  A more general control of the entanglement between photofragments  can be achieved by leveraging the interference between different multi-photon ionization (MPI) paths~\cite{Guillemin2015,ShapiroPRA2011}. In this latter approach, autoionizing resonances play a crucial role as intermediate states since they decay on a longer timescale than free photoelectron wavepackets~\cite{Argenti2010,ArgentiScience,EvaLuca_ChemRev,LucaPRL2014,Kotur2016,Ott2014}. In fact, metastable states are essential intermediates in resonant multi-photon atomic ionization \cite{Nagasono2007,Htten2018}, ultrafast electron decay \cite{Fohlisch2005} and molecular dissociative photoionization \cite{Sansone2010, Martin2007}.
\begin{figure}[hbtp!]
\begin{center}
\includegraphics[width=0.9\columnwidth]{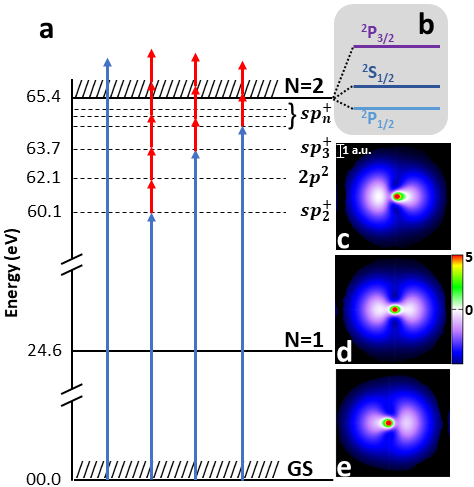}
\caption{\label{fig:energylevels}
a) An attosecond XUV pump pulse and an IR probe pulse cause the shake-up ionization of helium through several multi-photon paths, some of which involve intermediate autoionizing states. The Interference between direct and multi-photon ionization paths gives rise to a partial coherence between the $2s$ and $2p$ states of the ion, controllable via the pump-probe delay $\tau$. In the non-relativistic approximation, a $2s-2p$ coherence corresponds to a permanent polarization of the ion. c-e) Ion electron density at $\tau=0$, 1, and 2~fs; the light is polarized along the horizontal axis. b) Due to the fine-structure splitting of the $n=2$ He$^+$ level, the ionic dipole fluctuates, thus mapping the attosecond dependence of any initial ionic coherence to the picosecond timescale.
}
\end{center}
\end{figure}

In this work, we use \emph{ab initio} simulations to explore the control and reconstruction of the density matrix of the ensemble of $2s$ and $2p$ parent ions that emerge from the shake-up ionization of the helium atom with an XUV-pump IR-probe sequence of ultrashort pulses, linearly polarized along the $z$ axis~\cite{LucaAlvaroPRA2015,Ott2014,LucaPRL2014}. Multi-photon excitations are key to entangle the $2s$ and the $2p$ states, which have opposite parity. In a pump-probe ionization of helium, autoionizing states below the $N=2$ threshold are known to affect the branching ratio between shake-up channels~\cite{Argenti2010}, due to the interference between direct-ionization and resonant MPI paths. This same interference affects also the residual coherence between the $2s$ and $2p$ states of the He$^{+}$ ion. In the non-relativistic limit, these states are degenerate, and hence their coherence results in a permanent dipole moment. We demonstrate that the magnitude of the polarization can be controlled by changing the pump-probe delay. On a timescale of few picosecond, the dipole moment fluctuates even in absence of external fields, due to spin orbit interaction~\cite{bookSpringer}. Our results show how the slow dynamics of such polarized-ion ensemble can be controlled with attosecond precision. Conversely, from these fluctuations, it is possible to reconstruct the relative phase between the $2s_{m_s}$ and $2p_{m_l,m_s}$ states in the ion wavepacket at the time of its inception. In particular, this reconstruction gives access to the coherence between the $2s_{1/2}$ and the $2p_{1,-1/2}$ states, which is a sensitive probe of relativistic effects in attosecond ionization, since it vanishes only in the non-relativistic limit.

The paper is organized as follows. Section~\ref{sec:TheoreticalMethods} offers an overview of the \emph{ab initio} theoretical and numerical methods used to compute, at the end of a pulse sequence, the photoelectron distribution entangled with each ion.
Section~\ref{sec:Simulations} describes the pump-probe setup used for the simulations, it  discusses the partial photoelectron distributions as well as the corresponding reduced density matrix for the ion. Section~\ref{sec:Reconstruction} describes the reconstruction of the ionic coherence phase from the picosecond beating of the ion dipole. Finally, Section~\ref{sec:Conclusions} summarizes the conclusions and perspectives of this work.

\section{Theoretical Methods}\label{sec:TheoreticalMethods}

The axially symmetric ($M=0$) singlet states of helium are expanded in close-coupling (CC) functions of the form~\cite{Argenti2006,Argenti2010,ArgentiPRA2013}
\begin{equation}\label{eq:ccstate}
\begin{split}
\phi^{L}_{\alpha i}(\mathbf{x}_1,\mathbf{x}_2) &=\hat{\mathcal{A}}\,\theta(\zeta_1,\zeta_2)
\mathcal{Y}^{L0}_{L_a\ell_\alpha}(\Omega_1,\Omega_2)\frac{u_{N_aL_a}(r_1)}{r_1}\frac{f_{\ell_\alpha i}(r_2)}{r_2}.
\end{split}
\end{equation}
where $\hat{\mathcal{A}}$ is the antisymmetrizer, $x=(\vec{r},\zeta)$ are electronic spatial and spin coordinates, $\theta(\zeta_1,\zeta_2)$ is a singlet two-electron spin function, $\mathcal{Y}^{LM}_{L_a\ell_\alpha}(\Omega_1,\Omega_2)$ are bipolar spherical harmonics~\cite{Varshalovich}, $a$ is an ionic state with angular momentum $L_a$ principal quantum number $N_a$, $\ell_\alpha$ is a satellite-electron angular momentum, $u$ and $f$ are reduced hydrogenic and arbitary radial functions, respectively, and $\alpha=(L_a,N_a,\ell_\alpha,L)$ is a collective index that identifies a close-coupling channel. Unless specified otherwise, atomic units ($m_e=1$, $\hbar=1$, $e=1$) are used throughout.
To reproduce short-range correlation, the basis includes also a complementary set of states with the same expression as~\eqref{eq:ccstate}, with both $u$ and $f$ free to vary within an large set of localized functions. The reduced radial functions in Eq.~\eqref{eq:ccstate} are expanded in B-spline bases of order 10~\cite{deBoor,Bachau2001,Argenti2006}, reaching a maximum radius of $\sim 41$~a.u. and $R_{\textsc{BOX}}\sim 1200$~a.u. for the $u$ and $f$ functions, respectively.

The discretized eigenstates of the atom confined to a box with radius $R_{\textsc{BOX}}$ are obtained by diagonalizing the non-relativistic field-free Hamiltonian $H_0$,
\begin{equation}\label{eq:H0}
H_0=\frac{p_1^2}{2}+\frac{p_2^2}{2}-\frac{2}{r_1}-\frac{2}{r_1}+\frac{1}{r_{12}},
\end{equation} 
in the full close-coupling basis.
The multi-channel scattering states are obtained by solving the principal-part Lippmann-Schwinger equation (LSE)~\cite{Newton}, 
\begin{equation}\label{eq:LSE}
\psi_{\alpha E}^{\mathcal{P}} = \phi_{\alpha E} + \mathsf{G}_0^{\mathcal{P}}(E) V \psi_{\alpha E}^{\mathcal{P}},\quad \mathsf{G}_0^{\mathcal{P}}(E) \equiv \frac{\mathcal{P}}{E-\mathsf{H}_0},
\end{equation}
where $\mathsf{H}_0 = \sum_\alpha P_\alpha H_0 P_\alpha$ is a reference Hamiltonian in which the coupling between different channels are set to zero, $P_\alpha$ is the projector on a close-coupling channel, and $V=H_0-\mathsf{H}_0$ is the inter-channel coupling potential. Equation~\eqref{eq:LSE} is discretized and solved using the K-matrix method~\cite{Moccia1991,Mengali1996,Fang2000,Argenti2006,Argenti2007,Argenti2006R,Argenti2008b,Argenti2010b,EvaLuca_ChemRev,Argenti2016} which expresses LSE stationary solutions as
\begin{equation}
\psi_{\alpha E}^{\mathcal{P}}=\phi_{\alpha E}+\sum_\gamma\sumint d\epsilon\, 
\phi_{\gamma \epsilon}\frac{\mathcal{P}}{E-\epsilon}
\mathbf{K}_{\gamma\epsilon,\alpha E},
\label{eq:tm1}
\end{equation}
where $\mathbf{K}_{\gamma\epsilon,\alpha E}=\langle \phi_{\gamma \epsilon}|V|\psi^{\mathcal{P}}_{\alpha E}\rangle$, is the \emph{off shell} reactance matrix~\cite{Newton}. The scattering states with incoming boundary conditions $\psi_{a\vec{k}\sigma}$ are given by
\begin{equation}
\begin{split}
\psi_{a\vec{k}\sigma}^-&=\frac{(-)^{\frac{1}{2}-\sigma}}{\sqrt{k}}\sum_{L\ell m\beta} C_{L_aM_a,\ell m}^{L0}Y_{\ell m}^*(\hat{k})\psi_{\beta E}^{\mathcal{P}}\,\times\\
&\times\left[\frac{1}{\mathbf{1}-
    i\pi\mathbf{K}(E)}\right]_{\beta\alpha}e^{- i(\sigma_{\ell_\alpha}+\delta_\alpha-\ell_\alpha\pi/2)},
    \end{split}
\end{equation}
where $\vec{k}$ and $\sigma$ are the asymptotic photoelectron linear momentum and $z$ spin projection, respectively, $\mathbf{K}_{\alpha,\beta}(E)\equiv\mathbf{K}_{\alpha E, \beta E}$ is the \emph{on-shell} reactance matrix (\S $7.2.3$ in \cite{Newton}),  $\sigma_{\ell_\alpha}=\arg\left[\Gamma(\ell+1-i/k)\right]$ is the Coulomb phaseshift, and $\delta_\alpha$ is the channel phase shift.
\begin{figure*}[hbtp!]
\includegraphics[width=\textwidth]{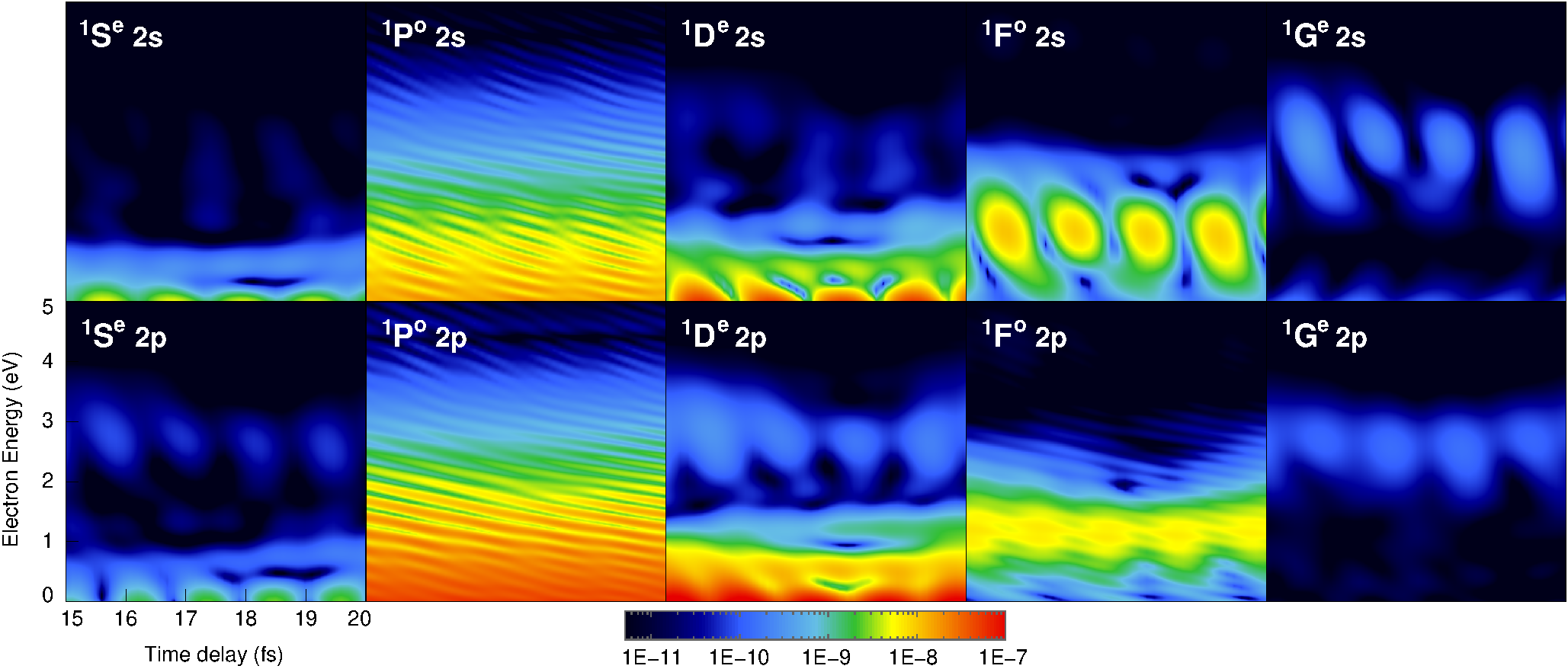}
\caption{\label{fig:dPaGdEvsT_V_cor} Symmetry-resolved partial differential photoelectron distributions, in velocity gauge, for the $2s$ channel (top panels) and $2p$ channel (bottom panels), as a function of the time delay and of the photoelectron energy. In the present simulation, the total spin is conserved and hence all the symmetries have singlet multiplicity. Since the light is linearly polarized, furthermore, all states are natural (same parity as the total angular momentum's) and $M_L=0$. The symmetries, therefore, are indexed by the total angular momentum $L$ only, which is shown here from $L=0$ (leftmost panel - $^1$S$^e$) to $L=4$ (rightmost panel - $^1$G$^e$).}
\end{figure*}

The time evolution of the atom in the presence of the external field is dictated by the time-dependent Schr\"odinger equation (TDSE),
\begin{equation}
i\partial_t\Psi(t) = \mathcal{H}(t) \Psi(t).
\end{equation}
where the total Hamiltonian $\mathcal{H}(t)=H_0 + H_I(t)$ comprises the time-dependent dipole interaction in velocity gauge $H_I(t)=\alpha \vec{A}(t)\cdot(\vec{p}_1+\vec{p}_2)$, where $\alpha = e^2/\hbar c\approx 1/137.036$~\cite{CODdataNIST} and $\vec{A}(t)$ is the transverse vector potential.
The TDSE is integrated in uniform time steps $dt$ 
\begin{equation}
\Psi(t+dt) = U_{\CAP}(dt)\,U(t+dt,t) \Psi(t),
\end{equation}
where $U(t+dt,t)$ is a second-order symmetrically split exponential unitary propagator,
\begin{equation}~\label{eq:unitaryPropagator}
U(t+dt,t)=e^{-i\,H\,dt/2}e^{-i\,H_I(t+dt/2) \,dt}e^{-i\,H \,dt/2},
\end{equation}
whereas $U_{\CAP}(dt)$ is an exponential evolution operator that accounts for the complex-absorption potential,
\begin{equation}~\label{eq:ucap}
U_{\CAP}(dt)=e^{-i\,dt V_{\CAP}},
\end{equation}
with $V_{\CAP}=-i C\sum_\alpha P_\alpha (r-R_\CAP)^2\theta(r-R_\CAP)$, and $C>0$, which prevents reflection at the boundary of the quantization box.
Partial photoelectron amplitudes are computed at the end of the pulse by projecting the wavefunction in interaction representation on a complete set of multi-channel scattering states for the two-electron system~\cite{Argenti2010,ArgentiPRA2013,Argenti2006R}, as a parametric function of the pump-probe delay $\tau$, $\mathcal{A}_{\alpha\vec{k}\sigma}(\tau)= \langle {\psi}_{\alpha, \vec{k} \sigma}^{-}|\Psi_I(t;\tau)\rangle$.

\section{Simulations}\label{sec:Simulations}
Figure~\ref{fig:energylevels}a illustrates the pump-probe excitation process we simulate. A weak single attosecond XUV pulse excites the neutral helium atom from the ground state to the $N=2$ shake-up ionization channels above the threshold, as well as to the DES below the $N=2$ ionization threshold. The $sp_{2}^{+}$ and  $sp_{3}^{+}$ states~\cite{Fano1961,ArgentiPRA2013,Rost_1997,RostPRL1996,JFiestPRL2011}, which are 5.04 eV and 1.69 eV below the $N=2$ threshold and have a lifetime of 17.6 fs and 80.3 fs, respectively, are populated most efficiently~\citep{Argenti2010}. The absorption of a single XUV photon cannot give rise to a coherent superposition of the $2s$ and $2p$ ionic states, since they have opposite parities and so have the photoelectron they are entangled with. To observe any coherence in the residual parent ion, therefore, it is necessary to associate the XUV pulse with additional control fields. In our simulation, an IR-probe pulse with a controllable delay with respect to the XUV pulse dresses the system at the time of the excitation and promotes the DESs to the shake-up ionization channels above the $N=2$ threshold. Thanks to the presence of several interfering multi-photon ionization-excitation paths, a coherence between degenerate $2s$, $2p$ ionic states does now emerge. 

The XUV pump pulse employed in the simulation has a Gaussian temporal profile, with central frequency $\hbar\omega_\XUV=60.69$~eV (2.2308~a.u.), a duration of 385~as (full width at half maximum of the envelope of the intensity, fwhm$_\XUV$), and a peak intensity $I_\XUV$=1~TW/cm$^2$. The IR probe pulse has a cosine-squared temporal profile, with central frequency $\hbar\omega_\IR=1.55$~eV (0.057~a.u.), an entire duration of 10.66~fs (fwhm$_\IR\approx$3.77~fs), and peak intensity $I_\IR=$1~TW/cm$^2$.
The electronic configuration basis comprises, beyond the minimal set of close-coupling channels $1s\epsilon_{\ell}$, $2s\epsilon_\ell$, and $2p\epsilon_\ell$, the full-CI set of configuration $n\ell n'\ell'$ constructed from all the localized orbitals with orbital angular momentum $\ell\leq 5$, and total angular momentum $L$ up to 9. The overall size of the {$^1$L$^\pi$} spaces, with $L=0$, $1$, $2$, $\ldots$, $9$, are: 9064, 12577, 13498, 12592, 12288, 11363, 10787, 10188, 9912, and 9672, for a total size of 111941. The energy of the ground state is $E_{\mathrm{g}}=-2.9036\,028$~a.u., to be compared with the accurate non-relativistic limit for $\ell_{\mathrm{max}}=5$, which is $-2.9036\,057$~a.u.~\cite{Carroll1979}.

The panels in Figure~\ref{fig:dPaGdEvsT_V_cor} show the partial photoelectron distributions for the $2s$ and $2p$ parent ions, resolved by total angular momentum, as a function of the pump-probe delay. The calculations are clearly converged with respect to the total angular momentum. Indeed, the largest angular momentum with an appreciable population is $L=4$ ({$^1$G$^e$}). Independent simulations carried out in length gauge produce virtually identical results, which gives further credence to the time-dependent calculations being well converged.
The reduced density matrix for the parent ion, $\rho_{\alpha\beta}$, is obtained tracing out the photoelectron states~\cite{FanoDM_1957},
\begin{equation}
\rho_{\alpha\beta}(\tau) = \sum_{\sigma}\int d^{3}k  \mathcal{A}_{\alpha\vec{k}\sigma}(\tau)\mathcal{A}^*_{\beta\vec{k}\sigma}(\tau).
\end{equation}
The coherence between ionic states~\cite{FanoDM_1957,Pabst2011} is defined here as 
\begin{equation}\label{eq:coherence}
    g_{\alpha\beta}(\tau) = \rho_{\alpha\beta}(\tau)/\sqrt{\rho_{\alpha\alpha}(\tau)\rho_{\beta\beta}(\tau)}
\end{equation}
\begin{figure}[hbtp!]
\begin{center}
\includegraphics[width=\columnwidth]{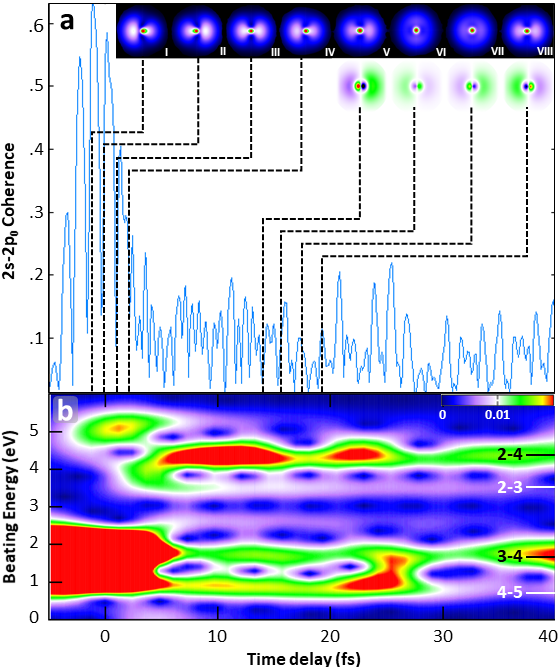}
\caption{\label{fig:Coherence}
a) Absolute value of the coherence between the $2s_{1/2}$ and $2p_{0,1/2}$ He$^+$ states, defined as in Eq.~\eqref{eq:coherence}, as a function of the pump-probe delay. In the region where two pulses overlap, the ion is polarized by the IR field, resulting in a large coherence with a time-delay period equal to the period of the IR field. Insets I-IV show the strong alignment of the ion electron density at the end of the pulse (the light is polarized along the horizontal axis). When the two pulses do not overlap, only multiphoton transitions that proceed through resonant states contribute to the shake-up ionization, and hence the time-delay dependence of the coherence is governed by the beating between DESs. In this case, the ion exhibits a smaller degree of coherence (insets V-VIII). b) Window Fourier Transform of the dipole moment, as a function of the time delay,  showing the transition from a single broad peak due to the ion polarization driven by the IR, when the pulses overlap, to multiple narrow peaks due to the beating between the $sp_n^+$ DESs, with $n=2-5$, which are the most populated.
}
\end{center}
\end{figure}

Figure~\ref{fig:energylevels}c-e show the ion electron density immediately after the ionization event, for a pump-probe delay $\tau$ of 0, 1, and 2~fs, respectively, computed from the \emph{ab initio} density matrix $\rho_{\alpha\beta}(\tau)$. The residual coherence results in a controllable polarization of the ion. 
Within the non-relativistic approximation, the $2s$ and $2p$ states are degenerate, and hence their dipole moment is stationary.  On the femtosecond timescale of the present simulation for helium, the non-relativistic approximation is expected to be valid. On longer timescales, however, relativistic interactions can no longer be neglected. Spin-orbit coupling splits the $2p$ level into a $^{2}$P$_{1/2}$ and $^{2}$P$_{3/2}$ multiplet~\cite{sakurai2017modern}, and Lamb shift lowers the energy of the {$^2$S$_{1/2}$} level compared to $^{2}$P$_{1/2}$~\cite{CODdataNIST,LambPRA}, see Figure~\ref{fig:energylevels}b. Due to these relativistic interactions, gathered in the fine-structure Hamiltonian $H_{fs}$, the density matrix undergoes periodic oscillations on a picosecond timescale, reproduced by the unitary transformation
\begin{equation}
\rho(t;\tau)=e^{-iH_{fs}t}\rho(\tau)e^{iH_{fs}t}.
\end{equation}
By the same token, the ion dipole moment is not stationary either, exhibiting fluctuations at the Bohr frequencies of the ion, $ \langle \mu_z(t;\tau)\rangle = Tr[\mu_z\rho(t;\tau)]$.
Figure~\ref{fig:Coherence}.a shows the absolute value of the coherence between the $2s$ and the $2p_z$ states as a function of the pump-probe delay. In the region where the two pulses overlap, ionization takes place in the presence of the IR probe pulse, which suppresses the channel in which the ion is polarized opposite to the IR field. As a consequence, the ion emerges strongly polarized, giving rise to the macroscopic polarization of the residual charge density shown in the insets of Figure~\ref{fig:energylevels}. The density fluctuates with the same frequency as the IR period, whereas coherence is maximum every half IR period, near the peak of the IR. 
When the two pulses do not overlap, beyond 4 fs time delay, the coherence exhibits weaker modulations due to the beating between the MPI amplitudes from the multiple intermediate doubly-excited states below threshold. The change in the charge density can be better appreciated from the left-right density asymmetry. Figure~\ref{fig:Coherence}.b shows the window Fourier transform of the dipole moment with respect to the time delay, 
\begin{equation}
\tilde{\mu}(\tau_w,\omega_\tau) = \frac{1}{\sqrt{8\pi^3}\sigma_w}\int d\tau\,e^{i\omega_\tau \tau-(\tau-\tau_w)^2/2\sigma^2_w} \mu(\tau),
\end{equation} 
where $\sigma_w=2.4$~fs, which features clear peaks, as a function of $\omega_\tau$, each corresponsing to the beating between a pair of doubly excited states. The spectrum is dominated by the beating between the pair of doubly excited states $sp_{2}^{+}$-$sp_{3}^{+}$, $sp_{2}^{+}$-$sp_{4}^{+}$, $sp_{3}^{+}$-$sp_{4}^{+}$, and $sp_{4}^{+}$-$sp_{5}^{+}$, which in the Figure are labelled $2-3$, $2-4$, $3-4$, and $4-5$, respectively. The beating with the $sp_{2}^{+}$ state decay more rapidly than the others, due to the short lifetime of this state ($\sim$18~fs). All peaks exhibit strong modulations as a function of the window central delay, which shows that the pump-probe delay can be used as a femtosecond knob to control the degree of coherence of the ion.

\begin{figure}[hbtp]
\begin{center}
\includegraphics[width=\columnwidth]{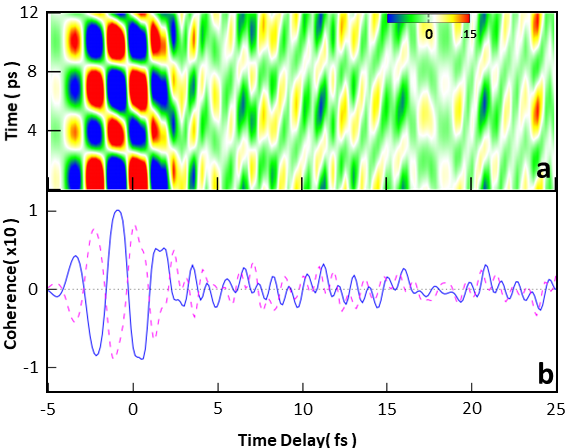}
\caption{\label{fig:dipole}
a) Polarized ions give rise to a dipole that oscillates as a function of the XUV-IR time delay on a timescale of few femtoseconds. In the non-relativistic limit, the 2s and 2p states of He$^{+}$ are degenerate, and hence, the dipole moment emerging from the pump-probe ionization process is permanent. However, the relative phase of the different J components of the ion does evolve in time, due to relativistic effects, resulting in the fluctuation of the dipole moment of the N=2 He$^{+}$ ion ensemble with a dominant period of $\sim$ 6 ps.
b) Real (blue solid line) and imaginary (purple dashed line) part of $\rho_{2s,2p_{0}}$, reconstructed from the periodic oscillation of the dipole on a picosecond timescale, which coincide with the actual quantities in the simulation.
}
\end{center}
\end{figure}

The splitting of the $n=2$ level causes the ionic dipole to oscillate in real-time, on a picosecond timescale. Figure~\ref{fig:dipole}a shows the ionic dipole as a function of both pump-probe delay and real time. When the pump and probe pulses overlap, the dipole fluctuates with a period of $\sim$ 6 ps, with its phase flipping periodically between 0 and $\pi$, giving rise to a checkerboard structure, as shown in Fig.~\ref{fig:dipole}a for time delays between $-5$ and $3$~fs. Since the $N=2$ manifold splits into three levels, the real-time beating contains two distinct frequencies, 24.5$\times$10$^{-6}$ a.u. and 2.1$\times$10$^{-6}$ a.u.~\cite{CODdataNIST}. From the picture, however, only the faster beating is clearly visible, since it is considerably stronger than the other. Furthermore, the longer period, $\sim 72$~ps, is close to a multiple of the faster period, of 6~ps, which reduces its visibility further. Nevertheless, from the Fourier Transform of the signal, both components can be accurately retrieved. The $6$~fs beating dominates the real-time evolution of the dipole even when the pump and probe pulses do not overlap. In contrast to the overlapping case, the phase of the oscillation now changes gradually as a function of the pump-probe delay. Indeed, in this time-delay range, the ion coherence originates from resonant multiphoton interferences, as shown by the Windowed Fourier Transform in Fig.~\ref{fig:Coherence}b, for time delays larger than $8$~fs. As a result, the relative phase of the DESs, which is encoded in the ion's permanent dipole shortly after the end of the pulse sequence, manifests itself in the femtosecond beating of the dipole as a function of the pump-probe delay as well as in the picosecond real-time oscillations of the dipole, stretched by three orders of magnitude.
Figure~\ref{fig:dipole}b shows the real and imaginary part of $\rho_{2s_\sigma, 2p_{0,\sigma}}$, computed at the end of pulse, which can both be retrieved from the long-time evolution of the dipole under the effect of fine-structure interactions.

\section{Reconstruction of coherence terms} \label{sec:Reconstruction}
The present excitation scheme has a duration of few tens of femtoseconds, i.e., two orders of magnitude smaller than the spin precession period caused by the fine-structure splitting. As long as the electron spin does not affect the excitation process, therefore, the dipole expectation value at the end of the pulses is dictated only by the coherence between the $2s_\sigma$ and $2p_{0,\sigma}$ states (the coherence is the same for $\sigma=\pm 1/2$), whereas the coherence between the $2s_\sigma$ and the $2p_{2\sigma,-\sigma}$ states is zero.  At larger times, the non-stationary character of the $2p_{m,\sigma}$ configurations emerges, and the dipole moment is observed to oscillate. When the fine-structure is taken into account, the time dependence of the dipole moment is dictated by two independent non-vanishing coherences, namely, those between the $|{^2S_{1/2, 1/2}}\rangle$ state and the two $|{^2P_{j 1/2}}\rangle$ states, for $j=1/2$ and $j=3/2$. These coherences beat at different frequencies, 
\begin{equation}\label{RE_DMZ}
  \langle \mu_{z}(\tau,t)\rangle \propto \sum_{j=\frac{1}{2},\frac{3}{2}} P_{j}(\tau) \cos(\omega_jt-\phi_j(\tau)),
\end{equation}
where $\omega_j=E_{^2P_{j}}-E_{^2S_{1/2}}$, and hence they can be separately measured.
Neglecting the small differences in their radial wave functions, we can write the $^{2}P_{1/2}$ and $^{2}P_{3/2}$ fine-structure states in terms of the $2p_{m,\sigma}$ spin orbitals just by coupling the orbital and spin angular momentum, $|{^2P_{j\mu}}\rangle=\sum_{m\sigma}|2p_{m,\sigma}\rangle C_{1m,\frac{1}{2}\sigma}^{j\mu}$, where $C_{a\alpha,b\beta}^{c\gamma}$ are Clebsch-Gordan coefficients~\cite{sakurai2017modern,bookSpringer}. The spin-free character of the ultrashort excitation process, therefore, causes the coherence between the $|{^2S_{1/2, 1/2}}\rangle$ and $|{^2P_{1/2, 1/2}}\rangle$ states to be in geometrically fixed proportion to the coherence between the $|{^2S_{1/2, 1/2}}\rangle$ and $|{^2P_{3/2, 1/2}}\rangle$ states. This circumstance allows us to predict, from the \emph{ab initio} spin-free attosecond pump-probe simulations, the time evolution of the dipole at large times. From the window Fourier transform $\mathcal{F}_t$ of the ionic dipole as a function of time $t$, we can obtain the phase and amplitude of the signal at any Bohr frequencies $\Omega$, as a parametric function of the time delay $\tau$
\begin{equation}\label{WFT2}
\mathcal{F}_t\left[\mu_{z}(t;\tau)w(t)\right](\Omega;\tau) \propto \sum_{j,\sigma=\pm} P_{j}(\tau) e^{i\sigma\phi_{j}(\tau)} w(\Omega-\sigma\omega_{j}),
\end{equation}
where $w(t)$ is a window function of time with fwhm much larger than the ion's Bohr beating periods, while $\sigma=\pm1$ corresponds to positive and negative frequencies, respectively. The FR in~\eqref{WFT2} exhibits isolated peaks at $\Omega = \omega_{j}$. The amplitude and the phase of any specific frequency, therefore, can be retrieved from the FT evaluated at that frequency, 
\begin{equation}\label{AMP_PHASE}
P_{j}(\tau)e^{\iota \phi_{j}} = C \frac{\mathcal{F}_t[\mu_{z}(t;\tau)w(t)](\omega_{j};\tau)}{w(0)},
\end{equation}
where $C$ is a constant common to all $j$s. In particular, it is possible to reconstruct the \emph{relative} phase between different beatings.
Conversely, from the phases and the relative amplitude of the dipolar beatings on the picosecond time scale, regardless if measured or simulated, it is possible to reconstruct the relative amplitude and phases of the coherences in the $\{2s_\sigma,2p_{m\sigma'}\}$ basis, at the end of the ultrashort sequence,
\begin{equation}\label{DM_NONZERO}
  \rho_{2s_\sigma,2p_{\sigma-\sigma',\sigma'}} = \sum_{j=\frac{1}{2},\frac{3}{2}} 
  C_{1 \sigma-\sigma',\frac{1}{2} \sigma'}^{j\sigma}
  \rho_{{^2\mathrm{S}}_{1/2,\sigma}, {^2\mathrm{P}}_{j,\sigma}},
\end{equation}
where we have used $m=\sigma-\sigma'$, since the z component of the total angular momentum (orbital plus spin) is conserved. The off-diagonal terms $\rho_{{^2\mathrm{S}}_{1/2,\sigma}, {^2\mathrm{P}}_{j,\sigma}}$ are related to the observable beating parameters, $\rho_{{^2\mathrm{S}}_{1/2,\sigma}, {^2\mathrm{P}}_{j,\sigma}} = P_j\,e^{i\phi_j}\,/\,\mu_{ {^2\mathrm{P}}_{j,\sigma},{^2\mathrm{S}}_{1/2,\sigma}}$.
To check the consistency of this reconstruction method, we have used it to recover the complex $\rho_{2s_\sigma,2p_{0,\sigma'}}$ coherences from the simulated long-time dipole fluctuation. The non-vanishing quantity $\rho_{2s_\sigma,2p_{0,\sigma}}$ so retrieved coincides with the one directly computed from the ionization wave function at the end of the pulse, which is plotted in Figure~\ref{fig:dipole}b.
Our \emph{ab initio} codes, which assume the non-relativistic approximation, predict the ratio $R=\rho_{2s_\sigma, 2p_{2\sigma,-\sigma}}/\rho_{2s_\sigma, 2p_{0,\sigma}}$ to be zero. As expected, our numerical reconstruction of this ratio from the asymptotic dipole beating reproduces this \emph{ab initio} prediction, which indicates the reconstruction procedure is accurate.
On the other hand, $R$ is not expected to vanish if the spin-orbit coupling or other fine-structure interaction has any role in the ultrafast ionization process. An experimental measurement of $R$, therefore, would open a new sensitive window on relativistic effects in attosecond ionization.

 Is it possible to gain experimental access to the relative amplitude and phase of the picosecond dipole beatings? In principle, the picosecond dipole beating can be measured using microwave spectroscopy~\cite{MVSPEC_PRA,MVSPEC_PRL}. The optical density of any the ionic ensemble generated by any realistic attosecond setup, however, is probably too small to be probed with microwave spectroscopic methods. A possible alternative way to measure the coherence of the $2s$ and $2p$ states is to map it to the population of the $N=3$ level by means of a combination of the 2$^{nd}$ and 3$^{rd}$ harmonics of the IR probe pulse, together with a delayed 5$^{th}$ harmonics. These transitions require a temporal resolution of about one picosecond, and hence their synchronization is not as challenging as the attosecond synchronization between the initial pump and probe pulses. By changing the delay between (2$^{nd}$+3$^{rd}$) and 5$^{th}$ harmonics, it is possible to change the total population transferred to the $N=3$ level in a predictable way. A last intense IR pulse can be used to fully ionize the $N=3$ states, whose population is finally measured by detecting the doubly-charged ion. The details of these possible experiments are beyond the scope of the present theoretical investigation.

\section{Conclusion}\label{sec:Conclusions}
In Conclusion, we have shown that the attosecond XUV-pump IR-probe ionization of helium can give rise to a coherence between the $2s$ and $2p$ ionic states, which can be controlled via the pump-probe delay, on a femtosecond time scale. When the pump and probe pulses overlap, the ionic coherence is due to the strong polarization of the ion within the field of the intense IR probe pulse. When the pump and probe pulses do not overlap, the ion still exhibits partial coherence thanks to the resonant quantum paths promoted by intermediate doubly-excited states. We demonstrate that the slow evolution of the dipole, due to the fine structure of the ion, maps on a picosecond time scale the relative DES phases. This slow evolution allows us to reconstruct the relative amplitude and phases of the ion coherences at the time of ionization. This reconstruction protocol not only gives access to the instantaneous polarization of the ion at its inception. It also offers a way to measure the coherence between states with anti-parallel spin projection at the time of the ionization, which quantifies the influcence of relativistic interactions on attosecond photoemission processes. 

\begin{acknowledgments}
This work is supported by NSF grant no. 1607588. E. Lindroth acknowledges support from Swedish Research Council, Grant No. 2016-03789. Special thanks to UCF Advanced Research and Computing Center for proving us with the facility of STOKES super computer. We would also like to thank Thomas Gallagher for useful discussions.
\end{acknowledgments}

\bibliography{PRL_REF_2019,bib2,PETSc}

\end{document}